\documentclass[aps,prb,twocolumn,groupedaddress,showpacs,preprintnumbers,amsmath,amssymb]{revtex4}
\usepackage[pdftex]{graphicx}
\usepackage{dcolumn}
\usepackage{bm}

\begin{document}


\title{Electron spin blockade and singlet-triplet transition in a silicon single electron transistor}


\author{Binhui Hu}
\author{C.H. Yang}
\affiliation{Department of Electrical and Computer Engineering, University of Maryland, College
Park, MD 20742}


\date{\today}

\begin{abstract}
We investigate a silicon single-electron transistor (SET) in a metal-oxide-semiconductor (MOS)
structure by applying a magnetic field perpendicular to the sample surface. The quantum dot is
defined electrostatically in a point contact channel and by the potential barriers from negatively
charged interface traps. The magnetic field dependence of the excitation spectrum is primarily
driven by the Zeeman effect. In the two-electron singlet-triplet (ST) transition, electron-electron
Coulomb interaction plays a significant role. The evolution of Coulomb blockade peaks with magnetic
field $B$ is also owing to the Zeeman splitting with no obvious orbital effect up to 9 T. The
filling pattern shows an alternate spin-up-spin-down sequence. The amplitude spectroscopy allows
for the observation of the spin blockade effect, where the two-electron system forms a singlet
state at low fields, and the spin polarized injection from the lead reduces the tunneling
conductance by a factor of 8. At a higher magnetic field, due to the ST transition, the spin
blockade effect is lifted and the conductance is fully recovered.
\end{abstract}

\pacs{73.23.Hk, 72.25.-b, 73.40.Qv}

\maketitle


\section{INTRODUCTION}

As a promising candidate of spin qubits for quantum computing, semiconductor quantum dots have
attracted tremendous research effort.\cite{dloss98} Recent progress in GaAs lateral quantum dots
has demonstrated all possible single-qubit operations\cite{Koppens06} and the square-root-of-swap
operation on two qubits,\cite{Petta05} which form a universal set for quantum
computing.\cite{dloss98} Compared to GaAs, lateral quantum dots in $^{28}$silicon are expected to
have a spin coherence time orders of magnitudes longer, because $^{28}$Si has no nuclear spin, and
there is no hyperfine interaction between electron spins and nuclear
spins.\cite{Feher59}\nocite{Tyry03} There has been steady progress in the development of
silicon-based single electron transistors using silicon-on-insulator (SOI) wafers,\cite{Fujiwara06}
Si/SiGe quantum well structures\cite{Simmons07} and metal-oxide-semiconductor (MOS)
structures,\cite{Jones06, Angus07} though it remains a considerable challenge due to material
properties.\cite{Jones06}

In this paper, we present our recent progress on an enhancement-mode MOS single electron transistor
(SET) in pure silicon.\cite{Jones06} The unique structure of the MOS-SET allows us to verify the
formation of quantum dots and better understand the device, which is largely neglected in some
early works. When properly biased, the investigated SET has a quantum dot in an electrostatically
defined point contact channel. In addition to single electron tunneling behavior, we have also
observed its magnetic field dependence. For all of the data presented here, the field is applied
perpendicular to the sample surface. The observed magnetic field dependence of the excitation
spectrum\cite{Kouwen01} is found to be primarily driven by the Zeeman splitting. Furthermore, the
spectrum enables us to directly observe the singlet-triplet (ST) transition, where
electron-electron Coulomb interaction plays a significant role. The evolution of Coulomb blockade
peaks with the magnetic field is also measured. The data strongly suggest that the ground state
energies also shift with the applied magnetic field owing to the Zeeman effect. Up to 9 T, there is
no obvious orbital effect. The evolution of peak amplitudes illustrates the spin blockade effect
due to magnetic field induced spatial separation of spin-up and spin-down states. At a higher
magnetic field, the spin blockade effect is lifted when the SET undergoes the ST transition.

\section{SAMPLES AND TRANSPORT CHARACTERISTICS }

\begin{figure}
\includegraphics[width=75mm]{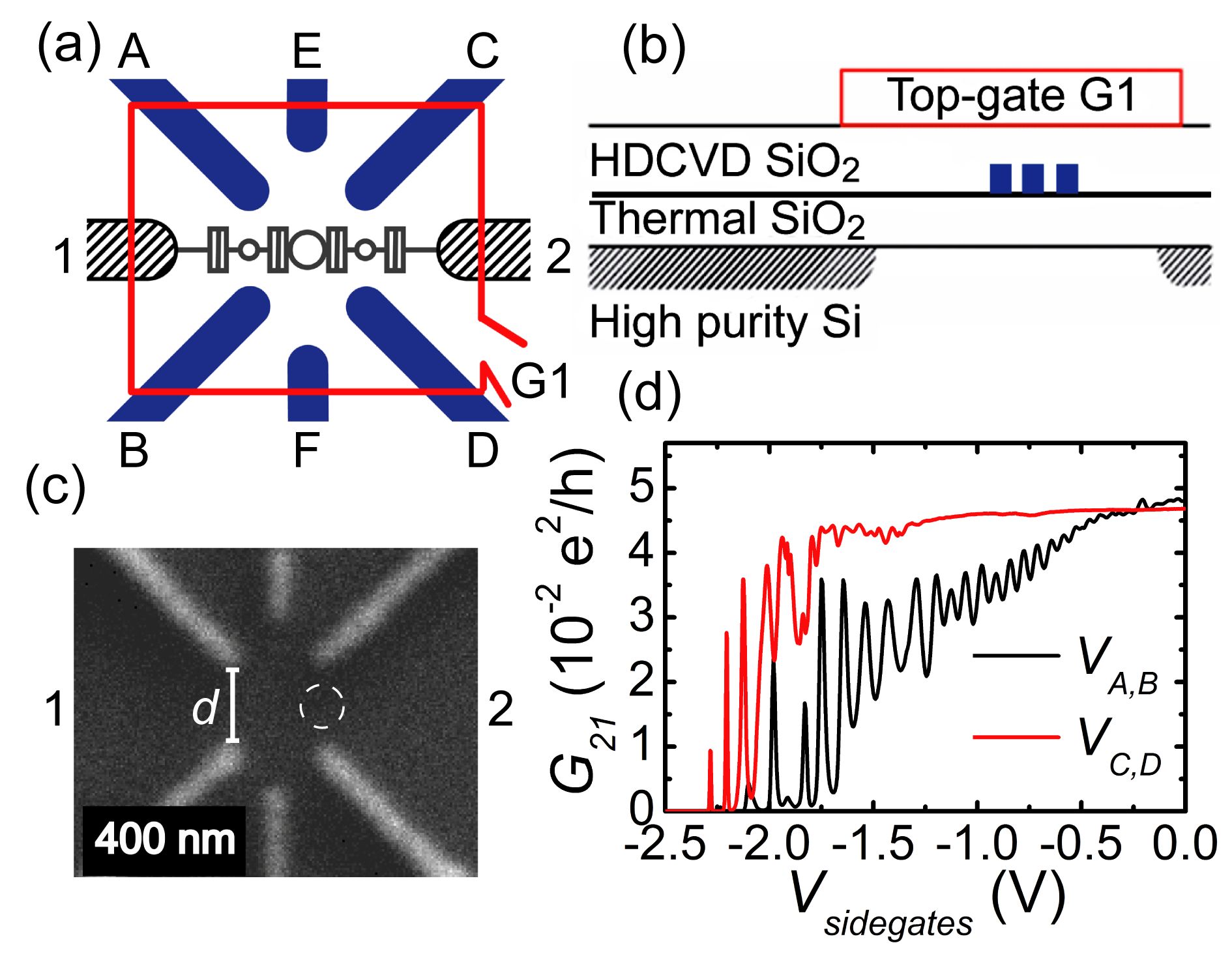}
\caption{\label{fig:schematics} (Color online) (a) Schematic top view and (b) cross sectional view
of a MOS-SET. In (a), the shaded regions, the solid lines, and the rectangular area, depict the
heavily phosphorus-doped ohmic source (1) and drain (2) leads, the 6 side gates (A-F), and the top
gate (G1) respectively. It also includes an equivalent circuit with two small quantum dots, due to
local potential fluctuation, and one large quantum dot at the center, which is electrostatically
defined by the 6 side gates. (c) Scanning electron micrograph of 70 nm-wide side gates before the
top SiO$_2$ layer is grown by high-density plasma-enhanced chemical vapor deposition (HDPECVD). The
gap, $d$, between two neighboring side gates is $\sim$160 nm. The circle depicts the location of
the quantum dot of interest as discussed in the text. (d) Transfer characteristics of single
quantum dots. The gate biases are: $V_{G1}=16$ V, $V_E=V_F=1$ V. In the first trace (depicted as
$V_{A,B}$), the side gate voltages $V_{A,B}$ ($=V_A=V_B$) are swept, when $V_C=V_D=0$ V. In the
second trace (depicted as $V_{C,D}$), the side gate voltages $V_{C,D}$ ($=V_C=V_D$) are swept with
$V_A=V_B=0$ V.}
\end{figure}

Figures \ref{fig:schematics}(a) and \ref{fig:schematics}(b) show the schematic top view and the
cross-sectional view of the device. All devices are fabricated on N-type, high purity silicon (100)
wafers with a resistivity of $3\sim5$ k$\Omega\cdot$m. By using photolithography and ion
implantation, the heavily phosphorous-doped regions are defined as the source (1) and the drain
(2). A 27 nm thick thermal oxide is grown by dry oxidation. It also has a bilayer-gated
configuration. Six side gates (labeled as A-F in Fig.\ \ref{fig:schematics}(a)) are located above
the thermal oxide and buried in the second dielectric layer, SiO$_2$, which is 400 nm in thickness
and grown by high-density plasma-enhanced chemical vapor deposition (HDPECVD). The top gate (G1),
which is above the HDPECVD oxide, laterally overlaps with the ohmic source and drain regions. The
device is annealed in forming gas at 420 $^\circ$C for 30 minutes to reduce the interface states.
The fabrication process and the basic operating principle have been discussed in more detail
elsewhere.\cite{Jones06} These devices are characterized in a dilution refrigerator with $\sim$10
mK base temperature. The electron temperature is about 400 mK by fitting the Coulomb blockade peaks
with the equation,\cite{Beena91}
\begin{equation}\label{eq:peak}
G/G_{max}=cosh^{-2}(e\alpha(V_g-V_{g0})/2k_BT),
\end{equation}
where $G$ is the source-drain conductance, $G_{max}$ is the maximum conductance at the peak,
$V_{g0}$ is the gate voltage at resonance, $k_B$ is the Boltzmann constant, and $T$ is the electron
temperature. The ratio of the gate capacitance to the total capacitance, $\alpha=C_g/C_\Sigma$, is
extracted from the slopes of nearby diamonds in the stability chart.\cite{Kouwen01} The
source-drain conductance is measured by the standard ac lock-in technique using a 37 Hz, 0.1 mV
excitation voltage. The positively biased top gate (G1) induces two-dimensional (2D) electrons at
the silicon and silicon thermal oxide interface. The peak electron mobility is about 5200 cm$^2$/Vs
at 400 mK in a test hall bar device, which has the same structure, but without the 6 side gates.
When $V_A=V_B=V_C=V_D=V_E = V_F = 0$ V, the SET device turns on at $V_{G1} > 9$ V. (Transport data
not shown here.) The negatively biased side gates A and B can deplete electrons below and define a
point contact channel. Due to the local potential fluctuation near the Si/SiO$_2$ interface, there
is a quantum dot formed in the channel, as illustrated in Fig.\ \ref{fig:schematics}(a). Side gates
C and D, and separately, gates E and F, can also be used to define a quantum dot through the same
mechanism. When the gates are properly biased, the device displays single quantum dot SET
characteristics. As shown in Fig.\ \ref{fig:schematics}(d), two typical data illustrate the
transfer behavior. In the first trace (depicted as $V_{A,B}$), $V_{G1} = 16$ V, $V_E = V_F = 1$ V,
$V_C = V_D = 0$ V, and side gates A and B ($V_A = V_B = V_{A,B}$) are biased from 0 to $-2.5$ V. As
$V_{A,B}$ is swept, the drain conductance oscillates; when $V_{A,B} < -2.2$ V, the conductance is
less than $1.3\times10^{-5}$  $(e^2/h)$, the noise floor of our measurement system. In the second
trace (depicted as $V_{C,D}$), the gate voltage $V_{C,D}$ ($=V_C=V_D$) is swept, while $V_A=V_B=0$
V, and the SET conductance also oscillates as $V_{C,D} < -1.3$ V and becomes diminished at $-2.5$
V. These pictures are well supported by the following data.

\begin{figure}
\includegraphics[width=75mm]{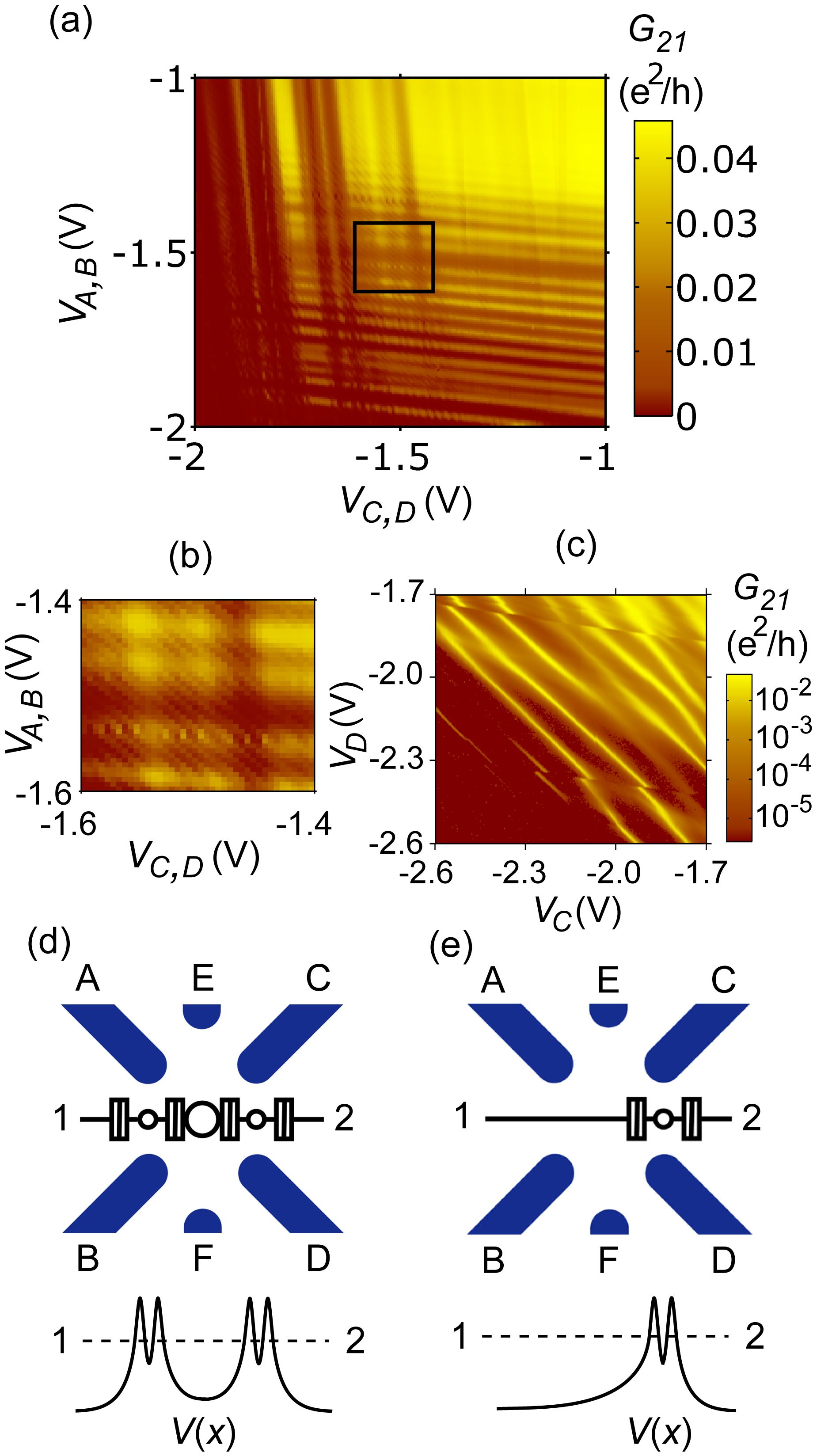}
\caption{\label{fig:dots}(Color online) Characteristics and equivalent circuits of the device at
$V_{G1}=16$ V, $V_E=V_F=1$ V. (a) Source-drain conductance $G_{21}$ as a function of side gate
voltages $V_{A,B}$ and $V_{C,D}$. (b) Enlarged section of (a) shows the underlining diagonal lines
from the center large quantum dot. (c) Source-drain  conductance $G_{21}$ as a function of $V_C$
and $V_D$ , with $V_A=V_B=0$ V. Diagonal lines suggest that side gate C and side gate D are coupled
to the small quantum dot with the same capacitance, as illustrated in (e). (d) Equivalent circuit
and schematic potential profile under multiple quantum dots bias conditions. (e) Equivalent circuit
and schematic potential profile under single quantum dot bias conditions.}
\end{figure}

To verify the formation of the quantum dots and to better understand the operation of the device,
we have investigated the dependence of the source-drain conductance $G_{21}$ on $V_{A,B}$ and
$V_{C,D}$. The observed transfer characteristic shown in Fig.\ \ref{fig:dots}(a), a portion of
which is enlarged and shown in Fig.\ \ref{fig:dots}(b), demonstrates a double quantum dot behavior.
When $V_{A,B}$ and $V_{C,D}$ are both less than about $-1.2$ V, the device is in the weakly coupled
double quantum dot regime. In this weakly coupled regime, two triple points will merge into ones,
i.e. the high conductance spots as clearly shown in Fig.\ \ref{fig:dots}(b), and are located at the
vertices of parallelograms.\cite{Van03} However, when $-1.2$ V $<V_{C,D} <$ 1 V, sweeping $V_{A,B}$
reproduces the single electron tunneling features similar to that shown in Fig.\
\ref{fig:schematics}(d). The same is observed when sweeping $V_{C,D}$ while keeping $-1.2$ V $<
V_{A,B} < 1$ V. The nearly horizontal and vertical lines in Fig.\ \ref{fig:dots}(a) indicate that
these two dots are spatially separated. One dot is strongly capacitively coupled to side gates A
and B, but only weakly coupled to gates C and D. The other, however, is strongly coupled to gates C
and D, and only weakly coupled to gates A and B. That is, one dot is physically near side gates A
and B, whereas the other dot is located close to side gates C and D. Furthermore, data shown in
Fig.\ \ref{fig:dots}(c) suggest that the quantum dot is located in the narrow point contact
channel. Keeping the other two point contacts fully conductive ($V_{G1}=16$ V, $V_E=V_F=1$ V, and
$V_A=V_B=0$ V) and sweeping the voltage of side gates C and D in the range of interest, the
source-drain conductance shows clear single electron tunneling characteristics. Because the
capacitances between the dot and, separately, side gates C and D are the same, the peak positions
in Fig.\ \ref{fig:dots}(c) display diagonal dependence.\cite{Van03} To be more specific, the
quantum dot is physically located in the point contact channel, at equal distances to side gates C
and D. The discontinuity of the diagonal lines is caused by a mere one electron charge change at a
nearby charge trap. Our experiment shows that this nearby charge trap has a negligible effect on
our results in this paper. A similar characteristic is also observed for the dot defined by side
gates A and B. (Data not shown here.) These observations are consistent with the picture that there
are two weakly-coupled quantum dots, located in the point contact channels defined by side gates A,
B, and C, D, respectively. Residue interface charges are known to form potential fluctuation at the
Si/SiO$_2$ interface.\cite{Kastner92} It has been reported earlier that a point contact channel in
silicon, due to these interface charges, can show Coulomb blockade
oscillations.\cite{Kastner92,Ishikuro97,MCLin07} In addition to the two quantum dots discussed
above, side gates A, B, C, D, E, and F, can also be biased to form a (large) quantum dot. A closer
look at Fig.\ \ref{fig:dots}(a) indeed reveals the underlining diagonal lines from this dot, as
shown in Fig.\ \ref{fig:dots}(b).\

In summary, under proper bias condition, $V_{A,B} <-1.2$ V and $V_{C,D} <-1.2$ V, the equivalent
circuit consists of an SET at the center (with a large quantum dot), in series with two others
(both with small quantum dots), as illustrated in Fig.\ \ref{fig:dots}(d); on the other hand, a
pair of side gates can define a single SET, when the other side gates are positively biased, as
shown in Fig.\ \ref{fig:dots}(e). These pictures do capture the major characteristics of the
device. However, there is not always a quantum dot in a point contact. When the gap between two
neighboring side gates is reduced from $\sim160$ nm to $\sim90$ nm, the point contacts can smoothly
turn off the device without Coulomb oscillations in some of our samples. In these devices, the
electrostatically defined dot is the only feature, but they currently only work in the
many-electron regime. With continuous down-scaling and improvement, it is quite possible to reach
the few-electron regime, even realize single electron confinement.

\section{SINGLE QUANTUM DOT AND  MAGNETIC FIELD SPECTROSCOPY}

\begin{figure}
\includegraphics[width=75mm]{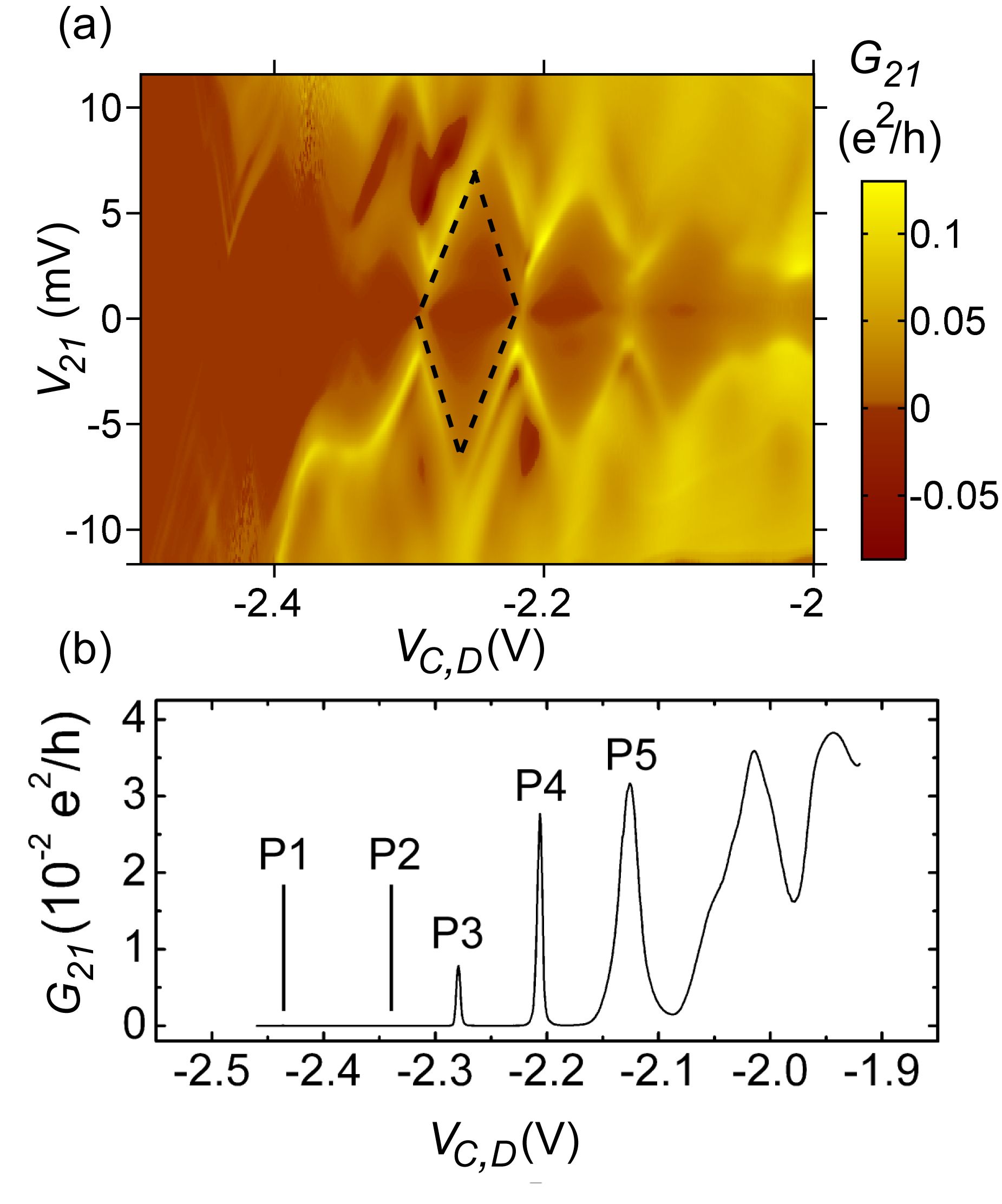}
\caption{\label{fig:diamond}(Color online) (a) Stability chart taken with $V_{G1}=16$ V,
$V_A=V_B=0$ V, $V_E=V_F=1$ V, and magnetic field $B=0$ T. (b) The Coulomb blockade oscillations.
The first 5 peaks, labeled as P1--P5, are shown. The data is taken under the same condition as that
in (a).}
\end{figure}

In the following, for the purpose of investigating single spin in silicon, we focus on the single
dot SET confined in the channel defined by side gates C and D. Fig.\ \ref{fig:diamond}(a) shows the
stability chart of the SET, where the source-drain differential conductance $G_{21}$ is measured
against the source-drain dc bias, $V_{21}$, and the side gate voltages $V_{C,D}$, while $V_E=V_F=1$
V and $V_A=V_B=0$ V. The observation of Coulomb diamonds confirms the formation of a single quantum
dot in the point contact channel. Within the orthodox theory, the quantum dot is modeled by a disc
with a diameter $d$, and the total capacitance $C_\Sigma$ is $4\varepsilon d$, where $\varepsilon$
($=11.9$ in silicon) is the dielectric constant. From the diamond shown in Fig.\
\ref{fig:diamond}(a), we can directly measure the half height ($V_{21}=e/C_\Sigma$), and the
obtained charging energy $E_c$ ($=e^2/C_\Sigma$) is about 6 meV. So $C_\Sigma$ is approximately 27
aF, which suggests a disc diameter of about 60 nm. This estimated disc diameter is of the same
order of magnitude as the averaged spacing between the interface-trapped charges in our device; we
have separately measured by the conductance method using large area devices the interface quality,
and obtained, at room temperature, an interface trap density of $\sim6\times10^{10}$
 eV$^{-1}$cm$^{-2}$, corresponding to a spacing between charges to be $\sim40$ nm.  The origin of the
quantum dot of interest is probably coming from negatively charged electron traps, located at the
Si/SiO$_2$ interface. Single dopants such as phosphorous atoms had been proposed to be behind
transport studies\cite{Sellier06} in silicon SETs, but our observed addition energy and the
excitation energy are both much less than what one would expect using the single dopants picture.
In addition, as will be discussed below, we have observed a spin filling pattern different from
that of the single phosphorous dopants picture. Other authors proposed that the quantum dots are
from Si nanoparticles in their point contact silicon-on-insulator (SOI) devices.\cite{Ishikuro97}
Our device is free of nanoparticles, thus different from them. Therefore, it is most likely that in
our system the negative interface trapped charges create potential barriers along the
one-dimensional (1D) point contact channel and form a quantum dot.\cite{Kastner92} These trapped
charges are always present in the Si/SiO$_2$, and the current industrial standard provides an
interface trap density of approximately $1\times10^{10}$ cm$^{-2}$. In all of the four SETs we have
characterized, the single electron tunneling characteristics are similar, although the details of
which are sensitive to the specific interface charge distribution. There are changes due to thermal
cycling, but the main features discussed here are stable and can be reproduced. Because the top
gate threshold voltage is measured to be $\sim 9$ V for the SET, using a parallel capacitor model
and that the quantum disc has a diameter of 60 nm, at $V_{G1}=16$ V the number of electrons in the
quantum dot is estimated to be at most $\sim10$. As the source-drain dc bias $V_{21}$ is fixed at 0
V, the SET displays the Coulomb blockade oscillations, as shown in Fig.\ \ref{fig:diamond}(b). The
following discussion will be focused on the magnetic field dependence of the first five peaks
labeled as P1--P5 in Fig.\ \ref{fig:diamond}(b).

\begin{figure}
\includegraphics[width=75mm]{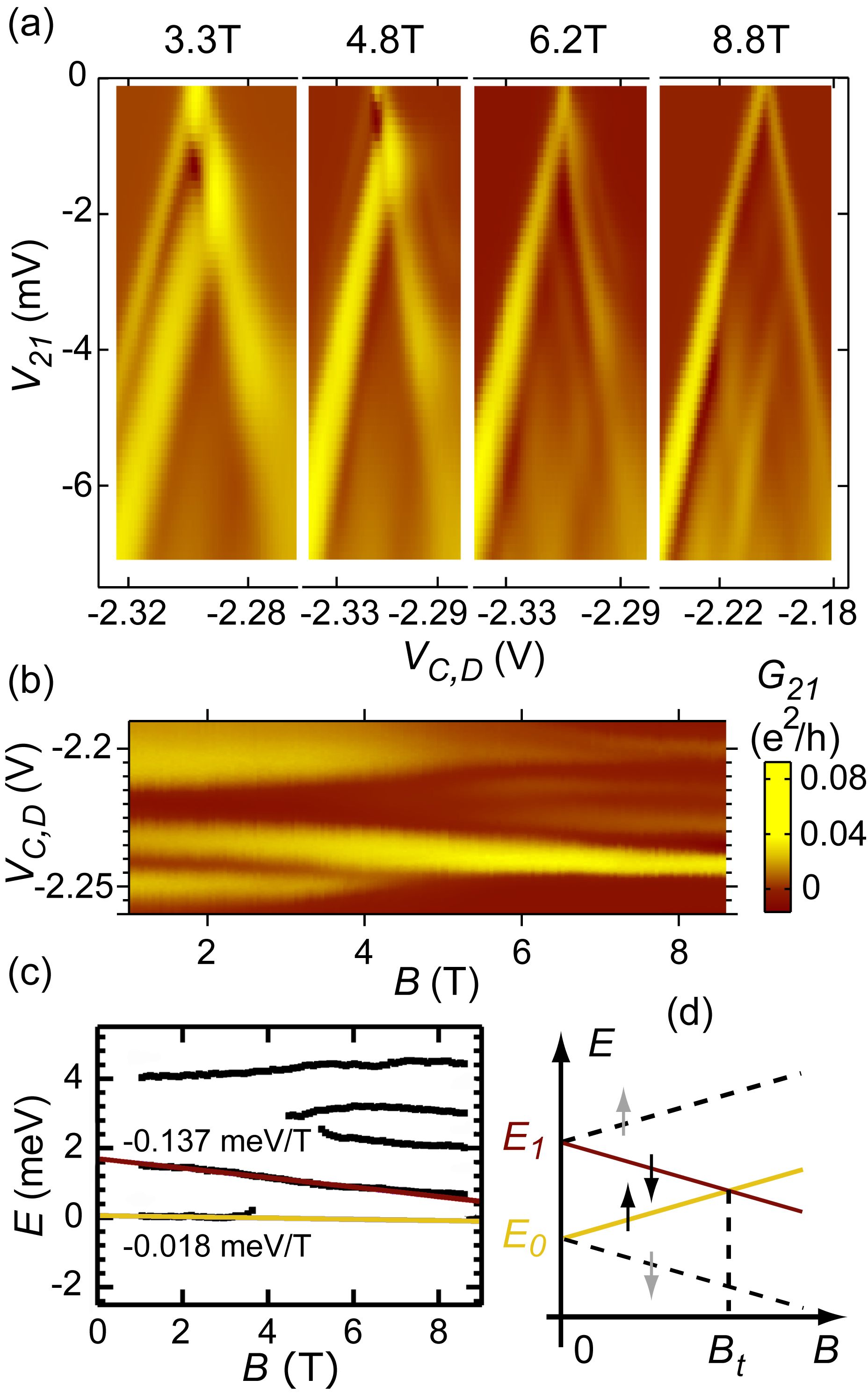}
\caption{\label{fig:stripe}(Color online) (a) Differential conductance $G_{21}$ as a function of
$V_{21}$ and the side gate voltage $V_{C,D}$ , near peak 4, at the magnetic fields of 3.3, 4.8,
6.2, and 8.8 T, respectively; $V_{G1}=16$ V, $V_A=V_B=0$ V, $V_E=V_F=1$ V. (b) The corresponding
excitation stripe taken at $V_{21}=-4.7$ mV with magnetic field $B$ between 1 and 9 T. (c) Peak
positions as a function of $B$ are extracted from the raw data in (b). The peak positions are in
chemical potential, $E=e\alpha V_{C,D}$, with an arbitrary offset. The straight lines are from
linear fitting of the data. The ground states $E_0$ are only fitted from 1 to 3 T. Their slopes are
labeled in the graph. (d) Schematic showing the evolution of single-particle energy levels $E_0$
and $E_1$ driven by the Zeeman effect, where $E_0$ is the ground state with a spin-up electron, and
$E_1$ is the first excited state with a spin-down electron, depicted in solid lines. In a
two-electron system, the $B$-evolution of the singlet and the triplet states follow $E_0(B)$ and
$E_1(B)$, respectively. Therefore we also label the singlet state as $E_0$ and the triplet state as
$E_1$. The crossing between $E_0$ and $E_1$, the singlet-triplet transition, occurs at $B_t$.}
\end{figure}

We first focus on the evolution of the ground state and the excited states near peak 4 in magnetic
field. The measured excitation spectrum is shown in Fig.\ \ref{fig:stripe}(b).\cite{Kouwen01} In
Fig.\ \ref{fig:diamond}(a), there is an excited state near peak 4 ($V_{C,D}$ is about $-2.2$ V).
Figure \ref{fig:stripe}(a) shows that this excited state moves toward the ground state as the
magnetic field increases. When the source-drain dc bias $V_{21}$ is fixed at $-4.7$ mV, and the
magnetic field and $V_{C,D}$ are swept, the excitation stripe near peak 4 is shown in Fig.\
\ref{fig:stripe}(b). With the increase of magnetic field ($0 < B < 6$ T), the distance between the
first excited state $E_1$ and the ground state $E_0$ decreases. As $B > 6$ T, the state $E_1$
becomes the new ground state. The maxima in Fig.\ \ref{fig:stripe}(b) are fitted with Gaussians (as
functions of $V_{C,D}$), and $V_{C,D}$ can be converted into chemical potential, using $E=e\alpha
V_{C,D}$, where $\alpha=0.09$, as defined in Eq.\ \eqref{eq:peak}.\cite{Kouwen01} After an
arbitrary offset, the resulting peak positions (in chemical potential) are shown in Fig.\
\ref{fig:stripe}(c). The absolute energy position of a peak in the excitation spectrum is
determined by practical experimental parameters; we therefore focus on the difference between the
most important features, that is, the energy difference between the ``ground state'' $E_0$ and the
``first excited state'' $E_1$. To first order, both $E_0$ and $E_1$ show apparent linear dependence
on magnetic field $B$. We therefore fit $E_0(B)$ (in the range 1 T $< B < 3$ T) and $E_1(B)$ (in
the range 1 T $< B < 9$ T) by linear lines. Straight lines for the guidance of the eye are shown in
Fig.\ \ref{fig:stripe}(c). The energy difference, $E_1(B)-E_0(B)$,
($=[-0.137-(-0.018)]$(meV/T)$\cdot B$), is $-0.119$ meV/T. The Zeeman splitting in bulk silicon is
0.116 meV/T ($=g\mu_BB$, with $g=2$, $\mu_B$ is the Bohr magneton.). It is clear that the energy
difference is dominated by the Zeeman effect. Figure \ref{fig:stripe}(d) illustrates the Zeeman
effect and the ST transition, where we only consider the two electrons in the outmost shell. The
lower lying electrons, if any, are ignored, for they do not influence the spin dynamics. For a
one-electron system, the ground state is a spin-down state in a magnetic field. The Zeeman
splitting has the linear magnetic field dependence, $\pm g\mu_BB/2$. In Fig.\ \ref{fig:stripe}(d),
the spin of the second electron is depicted by solid black arrows. When this second electron is
added to the quantum dot, it would interact with the electron already in the dot and form either
singlet or triplet states. As shown in Fig.\ \ref{fig:stripe}(d), quantum selection rule dictates
the spin of incoming electrons, that is, a spin-up state for the singlet state $E_0$, and a
spin-down state for the triplet state $E_1$. At low magnetic fields, $0 < B < B_t$, the singlet
state has a lower energy than that of the triplet state. While at a higher magnetic field
($B>B_t$), the triplet state $E_1$ becomes the ground state. The spin configuration of the ground
states of the two-electron system is therefore controlled by the magnetic field. Were the Zeeman
splitting the only effect to be considered, the magnetic field induced ST transition should occur
at about 14 T. This estimate is based on the data shown in Fig.\ \ref{fig:stripe}(c), and that
there is $\sim 1.6$ meV energy difference between $E_1$ and $E_0$ at zero field. When the size of
the electron wave functions shrinks for increasing magnetic field $B$, the interdependence of
Coulomb interaction and single-particle states becomes important. Here we only consider the
first-order corrections due to the electron-electron interaction in this two-electron system. The
chemical potential is $\mu_S(B)=E'_0+g\mu_BB/2+C_{00}(B)=E_0(B)+\Delta C_{00}(B)$ for the singlet
state, where $C_{00}(B)$ is the direct Coulomb interaction when both electrons are in the gound
state, $E'_0+C_{00}(0)=E_0(0)$, and $\Delta C_{00}(B)=C_{00}(B)-C_{00}(0)$. For the triplet state,
the chemical potential is $\mu_T(B)=E'_1-g\mu_BB/2+C_{01}(B)-|K_{01}(B)|=E_1(B)+\Delta
C_{01}(B)-\Delta |K_{01}(B)|$, where $C_{01}(B)$ and $K_{01}(B)$ are the direct Coulomb interaction
and the exchange interaction respectively when one electron occupies the ground state and the other
occupies the first excited state, $E'_1+C_{01}(0)-|K_{01}(0)|=E_1(0)$, $\Delta
C_{01}(B)=C_{01}(B)-C_{01}(0)$ and $\Delta |K_{01}(B)|=|K_{01}(B)|-|K_{01}(0)|$.\cite{Kouwen01}
Both the direct Coulomb interactions and the exchange interaction increase with increasing $B$,
since the size of the wave functions decreases. For the triplet state, we hypothesize that $\Delta
C_{01}(B)$ is largely compensated by $\Delta |K_{01}(B)|$, due to the apparent linear
$B$-dependence of the triplet state; while the singlet chemical potential $\mu_S(B)$ increases with
increasing B. So the electron-electron Coulomb interaction (including the direct Coulomb
interaction and the exchange interaction) plays a significant role here and drives the
singlet-triplet transition at a much lower magnetic field between 4 and 6 T. There are some minor
deviations from the linear $B$-dependence of the triplet state, which could be from other effects
in addition to this simplified picture. A detailed analysis requires the precise information of the
electrostatic potential in the quantum dot, which is beyond the scope of this paper. The
conductance of the singlet state drops when $B > 4$ T, and it becomes invisible after the ST
transition. This is due to the transport blockade effect, which will be discussed later.

\begin{figure}
\includegraphics[width=75mm]{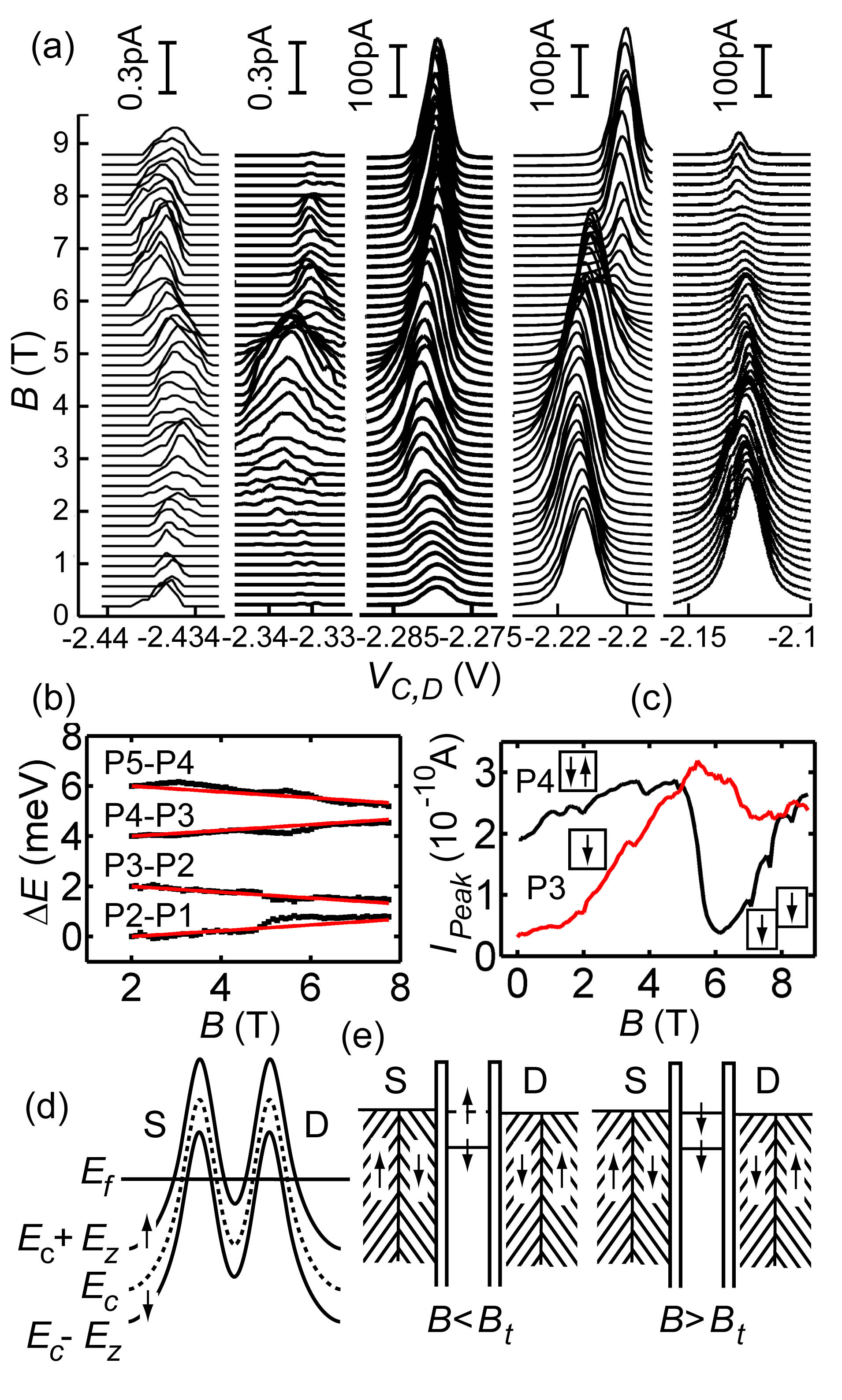}
\caption{\label{fig:peaks}(Color online) Evolution of Coulomb blockade peaks with the magnetic
field $B$. (a) Drain current of the first 5 peaks (P1--P5) with $V_{ac}=0.1$ mV, $V_{G1}=16$ V,
$V_A=V_B=0$ V, $V_E=V_F=1$. Each trace is offset linearly with $B$. (b) Peak spacing between
successive peaks and (c) peak amplitude as a function of $B$ are extracted from the raw data in (a)
fitted with Eq.\ \eqref{eq:peak}. In (b), the peak spacing is in chemical potential and is offset
for clarity. The straight lines have slopes $g\mu_BB$ or $-g\mu_BB$, assuming the Zeeman splitting
with $g=2$. In (c), the arrows in the squares indicate the spin configurations for peak 3 (P3) and
peak 4 (P4). (d) Schematic of the magnetic field induced spatial separation of spin-up and
spin-down states. In a magnetic field, the conduction band minima of spin-down and spin-up
electrons will split due to the Zeeman effect with $E_z= g\mu_BB/2$. Electrons are fully spin-down
polarized in the source and the drain leads near the quantum dot. (e) Schematic showing the
transport for a two-electron system with spin-down polarized leads. When $B<B_t$, the ground state
is a singlet, which only allows a spin-up electron to tunnel through, i.e., the spin blockade
state. When $B>B_t$, the ground state becomes a triplet, which permits a spin-down electron to
tunneling through. The spin blockade is thus lifted.}
\end{figure}

The magnetic field dependence of the first five Coulomb blockade peaks is shown in Fig.\
\ref{fig:peaks}(a). The peak position and the peak amplitude are extracted from this raw data by
fitting each peak with Eq.\ \eqref{eq:peak}. The shift of peak positions (in $V_{C,D}$) can be
converted into the change in chemical potential using the formula:$\Delta E(B)=e\alpha
[V_{C,D}^{Peak}(B)-V_{C,D}^{Peak}(0)]$. In the stability chart [Fig.\ \ref{fig:diamond}(a)], from
the slopes of Coulomb diamonds, $\alpha$ is about 0.15, 0.088, 0.090, 0.090 and 0.061 for peaks 1
to 5, respectively.\cite{Kouwen01} These changes in chemical potential reflect the evolution of the
ground state energies. Figure \ref{fig:peaks}(b) shows the peak spacing (in chemical potential)
between successive peaks as a function of magnetic field. Using the energy difference in our
analysis not only minimizes the uncertainty due to charge fluctuation and the long term drift in
analog electronics, but also helps in identifying the role of spin in the addition energy spectrum.
Based on the Zeeman splitting, four straight lines with slopes of either $g\mu_BB$ or $-g\mu_BB$,
with g=2, are plotted. Since the second peak only appears between 2 T and 8 T, we compare the first
5 peaks in this field range. Each curve is arbitrarily offset for clarity. Because the chemical
potential differences between successive peaks show the alternate slopes of $g\mu_B$ and $-g\mu_B$,
it is clear that the shifts of the ground states show an spin-down spin-up filling pattern and are
dominated by the Zeeman effect at low magnetic field ($B<4$ T). Note that such odd-even-odd-even
alternate filling pattern is different from that due to isolated single dopants in
silicon.\cite{Sellier06} The fact that the data are well explained by the Zeeman splitting is
consistent with our earlier finding that the magnetic field dependence of energy levels (both the
ground state and the excited state) is primarily driven by the Zeeman effect. For the data reported
in this work, the magnetic field is applied perpendicular to the sample surface, therefore, the
orbital effect might be expected. However, all of the observed magnetic field dependence can be
well explained by the Zeeman effect with a g-factor of $\sim 2$, leaving orbital effect not a
significant factor in our system. This is caused by the specific shape of our quantum dot. When we
squeeze side gates C and D, the electron wave function is elongated along the source-drain
direction, the orbital effect becomes negligible. This leaves the $B$ dependence of the peak
positions dominated by spin effects.\cite{Rohkin01}

The SET is fabricated on a Si (100) wafer, so the valley splitting from the two-fold valley
degeneracy is expected. It is generally believed in a Si (100) inversion layer that the valley
splitting ranges from $0 \sim 1.5$ meV, increases with increasing electric field or magnetic field,
and also depends on the Si/barrier interface.\cite{Ando82,Srijit07} But detailed relationships
between them are not clear yet. The valley splitting in Si quantum dots has not been systematically
investigated in experiments, primarily due to the lack of the excitation spectrum data. Some
speculate that it is $0.35\sim 0.46$ meV.\cite{Srijit07} Others indicate that it is of order of a
few meV and greater even at $B = 0$ T.\cite{Taka06,Liu08} In our SET, the spin-down spin-up filling
pattern requires that the valley degeneracy is lifted,\cite{Rohkin01} and the excitation spectrum
suggests that the valley splitting should be larger than 1.6 meV. This is probably due to the
strong lateral confinement in our quantum dot. Further investigation is necessary to reach an
affirmative conclusion.

Finally, we turn our attention to the tunneling peak amplitudes in Fig.\ \ref{fig:peaks}(a). In a
lateral quantum dot device, an electron tunnels into and out of the quantum dot from the
two-dimensional electron gas (2DEG) leads. As illustrated in Fig.\ \ref{fig:peaks}(d), the electron
concentration will vary from the bulk value to zero near the tunneling barriers; the Zeeman
splitting effect will cause the separation of the conduction band minima of the spin-up and
spin-down electrons in a magnetic field, which will result in the different populations of the
spin-up and spin-down subbands. In GaAs, spin polarized injection is due to edge states in a
magnetic field perpendicular to the 2DEG.\cite{Ciorga00} This edge-state picture is probably not
valid in our silicon devices, because the Shubnikov-de Hass oscillation minima are not zero up to 9
T in the test hall bar device. However, magnetic field can fully spin polarize a 2DEG with
$n_{2d}=1.5\times10^{11}$ cm$^{-2}$ for $B=5$ T solely with spin effects in a high mobility Si
MOSFET.\cite{Vitka01,Shash01} In the test hall bar device, the Shubnikov-de Hass oscillations show
that the 2DEG is spin polarized at least at $B=9$ T for $n_{2d}=7.12\times10^{11}$ cm$^{-2}$. So it
is reasonable to expect that electrons will be polarized into the spin-down state in the 2DEG leads
near the tunneling barriers above some magnetic field, as illustrated in Fig.\ \ref{fig:peaks}(d).
This effectively makes the spin-up electrons tunnel through a thicker barrier, so the injection is
dominated by spin-down electrons.

In theory, peak amplitude is proportional to the tunneling probability and depends exponentially on
how much wave functions in the dot and in the contacts overlap with each other.\cite{Rohkin01}
There are two different mechanisms governing the amplitude modulation of Coulomb blockade peaks in
a magnetic field. One is the spatial mechanism when an electron tunnels into different spatial wave
functions in a quantum dot, not related to spin; the other is the spin-blockade mechanism due to
spin-polarized injection.\cite{Sach01} The lower edge of the excitation stripe in Fig.\
\ref{fig:stripe}(b) reflects the evolution of peak 4 as a function of magnetic field in Fig.\
\ref{fig:peaks}(a). The conductance of the singlet state drops dramatically when $B>4$ T, while the
conductance of the triplet state doesn't change much, as shown in Fig.\ \ref{fig:stripe}(b) and
Fig.\ \ref{fig:peaks}(a). This difference can be explained by both the spatial effect and the
spin-blockade effect, since the singlet state requires a spin-up electron which occupies a smaller
wave function at the center, while the triplet state requires a spin-down electron which occupies a
larger wave function. However, the electron states corresponding to peak 3 (a spin-down single
electron state) and peak 4 (a singlet state) should have similar (single-particle) spatial wave
functions; their peak amplitude should have similar dependence on the magnetic field if there is no
spin blockade effect. In Fig.\ \ref{fig:peaks}(c), the amplitude of peak 4 decreases much faster
than peak 3 when the magnetic field is larger than 4 T. A reasonable explanation is the spin
polarized injection from the leads.\cite{Sach01} For 0.1 mV small ac excitation between the source
and the drain, only ground states can lie in the transport window. For a singlet state, only a
spin-up electron can tunnel into and out of the quantum dot, the amplitude of peak 4 decreases
dramatically owing to this spin polarized injection, and as shown in Fig.\ \ref{fig:peaks}(c), it
occurs at $B>5$ T. This forms the spin blockade. After $B>6$ T, the amplitude of peak 4 increases
with the magnetic field, because the ground state corresponding to peak 4 changes from a singlet
state to a triplet state. For the triplet state, the incoming spin-down electron experiences lower
and thinner barriers, thus the higher conductance. Since the amplitude of Coulomb blockade peaks is
determined by the two-electron spin configuration in the dot with spin polarized leads, it can be
used to distinguish a singlet state from a triplet one. Monitoring the amplitude is thus a form of
spin blockade spectroscopy.\cite{Ciorga00}

\section{CONCLUSIONS}

We have investigated electron spin in a nanometer-scale quantum dot in silicon. Using the unique
bilayer-gated structure, narrow point contact channels are electrostatically defined. Along the
narrow channel, a quantum dot is formed by potential barriers, probably due to negative charges
near the Si/SiO$_2$ interface. Tunneling spectroscopy clearly demonstrates the single electron
tunneling characteristics. There are two important spin-related features. First, applying
excitation spectroscopy on a dot populated by even number of electrons, we have directly observed
the ST transition. Second, based on the observed ST transition, we further use amplitude
spectroscopy, i.e., tracing the magnetic field dependence of Coulomb blockade peaks, to identify
the spin blockade phenomena. With a spin polarized injection, we demonstrate that the amplitude
spectroscopy can be used to detect a single spin.

\begin{acknowledgments}

This work is supported in part by the Laboratory for Physical Sciences and the Office of Naval
Research. The authors are grateful to Bruce Kane, Luyan Sun, Tomasz Kott, and Greg Jones, for
insightful discussions and technical help. The authors also acknowledge the Laboratory for Physical
Sciences for the use of their fabrication facilities.

\end{acknowledgments}
\bibliography{spinblockadeinSiSET}

\begin{thebibliography}{25}
\expandafter\ifx\csname natexlab\endcsname\relax\def\natexlab#1{#1}\fi
\expandafter\ifx\csname bibnamefont\endcsname\relax
  \def\bibnamefont#1{#1}\fi
\expandafter\ifx\csname bibfnamefont\endcsname\relax
  \def\bibfnamefont#1{#1}\fi
\expandafter\ifx\csname citenamefont\endcsname\relax
  \def\citenamefont#1{#1}\fi
\expandafter\ifx\csname url\endcsname\relax
  \def\url#1{\texttt{#1}}\fi
\expandafter\ifx\csname urlprefix\endcsname\relax\def\urlprefix{URL }\fi
\providecommand{\bibinfo}[2]{#2}
\providecommand{\eprint}[2][]{\url{#2}}

\bibitem[{\citenamefont{Loss and DiVincenzo}(1998)}]{dloss98}
\bibinfo{author}{\bibfnamefont{D.}~\bibnamefont{Loss}} \bibnamefont{and}
  \bibinfo{author}{\bibfnamefont{D.~P.} \bibnamefont{DiVincenzo}},
  \bibinfo{journal}{Phys. Rev. A} \textbf{\bibinfo{volume}{57}},
  \bibinfo{pages}{120} (\bibinfo{year}{1998}).

\bibitem[{\citenamefont{Koppens et~al.}(2006)\citenamefont{Koppens, Buizert,
  Tielrooij, Vink, Nowack, Meunier, Kouwenhoven, and Vandersypen}}]{Koppens06}
\bibinfo{author}{\bibfnamefont{F.~H.~L.} \bibnamefont{Koppens}},
  \bibinfo{author}{\bibfnamefont{C.}~\bibnamefont{Buizert}},
  \bibinfo{author}{\bibfnamefont{K.~J.} \bibnamefont{Tielrooij}},
  \bibinfo{author}{\bibfnamefont{I.~T.} \bibnamefont{Vink}},
  \bibinfo{author}{\bibfnamefont{K.~C.} \bibnamefont{Nowack}},
  \bibinfo{author}{\bibfnamefont{T.}~\bibnamefont{Meunier}},
  \bibinfo{author}{\bibfnamefont{L.~P.} \bibnamefont{Kouwenhoven}},
  \bibnamefont{and} \bibinfo{author}{\bibfnamefont{L.~M.~K.}
  \bibnamefont{Vandersypen}}, \bibinfo{journal}{Nature (London)}
  \textbf{\bibinfo{volume}{442}}, \bibinfo{pages}{766} (\bibinfo{year}{2006}).

\bibitem[{\citenamefont{Petta et~al.}(2005)\citenamefont{Petta, Johnson,
  Taylor, Laird, Yacoby, Lukin, Marcus, Hanson, and Gossard}}]{Petta05}
\bibinfo{author}{\bibfnamefont{J.~R.} \bibnamefont{Petta}},
  \bibinfo{author}{\bibfnamefont{A.~C.} \bibnamefont{Johnson}},
  \bibinfo{author}{\bibfnamefont{J.~M.} \bibnamefont{Taylor}},
  \bibinfo{author}{\bibfnamefont{E.~A.} \bibnamefont{Laird}},
  \bibinfo{author}{\bibfnamefont{A.}~\bibnamefont{Yacoby}},
  \bibinfo{author}{\bibfnamefont{M.~D.} \bibnamefont{Lukin}},
  \bibinfo{author}{\bibfnamefont{C.~M.} \bibnamefont{Marcus}},
  \bibinfo{author}{\bibfnamefont{M.~P.} \bibnamefont{Hanson}},
  \bibnamefont{and} \bibinfo{author}{\bibfnamefont{A.~C.}
  \bibnamefont{Gossard}}, \bibinfo{journal}{Science}
  \textbf{\bibinfo{volume}{309}}, \bibinfo{pages}{2180} (\bibinfo{year}{2005}).

\bibitem[{\citenamefont{Feher}(1959)}]{Feher59}
\bibinfo{author}{\bibfnamefont{G.}~\bibnamefont{Feher}},
  \bibinfo{journal}{Phys. Rev.} \textbf{\bibinfo{volume}{114}},
  \bibinfo{pages}{1219} (\bibinfo{year}{1959}); \bibinfo{author}{\bibfnamefont{A.~M.} \bibnamefont{Tyryshkin}},
  \bibinfo{author}{\bibfnamefont{S.~A.} \bibnamefont{Lyon}},
  \bibinfo{author}{\bibfnamefont{A.~V.} \bibnamefont{Astashkin}},
  \bibnamefont{and} \bibinfo{author}{\bibfnamefont{A.~M.}
  \bibnamefont{Raitsimring}}, \bibinfo{journal}{Phys. Rev. B}
  \textbf{\bibinfo{volume}{68}}, \bibinfo{pages}{193207}
  (\bibinfo{year}{2003}).

\bibitem[{\citenamefont{Fujiwara et~al.}(2006)\citenamefont{Fujiwara, Inokawa,
  Yamazaki, Namatsu, Takahashi, Zimmerman, and Martin}}]{Fujiwara06}
\bibinfo{author}{\bibfnamefont{A.}~\bibnamefont{Fujiwara}},
  \bibinfo{author}{\bibfnamefont{H.}~\bibnamefont{Inokawa}},
  \bibinfo{author}{\bibfnamefont{K.}~\bibnamefont{Yamazaki}},
  \bibinfo{author}{\bibfnamefont{H.}~\bibnamefont{Namatsu}},
  \bibinfo{author}{\bibfnamefont{Y.}~\bibnamefont{Takahashi}},
  \bibinfo{author}{\bibfnamefont{N.~M.} \bibnamefont{Zimmerman}},
  \bibnamefont{and} \bibinfo{author}{\bibfnamefont{S.~B.}
  \bibnamefont{Martin}}, \bibinfo{journal}{Appl. Phys. Lett.}
  \textbf{\bibinfo{volume}{88}}, \bibinfo{pages}{053121}
  (\bibinfo{year}{2006}).

\bibitem[{\citenamefont{Simmons et~al.}(2007)\citenamefont{Simmons, Thalakulam,
  Shaji, Klein, Qin, Blick, Savage, Lagally, Coppersmith, and
  Eriksson}}]{Simmons07}
\bibinfo{author}{\bibfnamefont{C.~B.} \bibnamefont{Simmons}},
  \bibinfo{author}{\bibfnamefont{M.}~\bibnamefont{Thalakulam}},
  \bibinfo{author}{\bibfnamefont{N.}~\bibnamefont{Shaji}},
  \bibinfo{author}{\bibfnamefont{L.~J.} \bibnamefont{Klein}},
  \bibinfo{author}{\bibfnamefont{H.}~\bibnamefont{Qin}},
  \bibinfo{author}{\bibfnamefont{R.~H.} \bibnamefont{Blick}},
  \bibinfo{author}{\bibfnamefont{D.~E.} \bibnamefont{Savage}},
  \bibinfo{author}{\bibfnamefont{M.~G.} \bibnamefont{Lagally}},
  \bibinfo{author}{\bibfnamefont{S.~N.} \bibnamefont{Coppersmith}},
  \bibnamefont{and} \bibinfo{author}{\bibfnamefont{M.~A.}
  \bibnamefont{Eriksson}}, \bibinfo{journal}{Appl. Phys. Lett.}
  \textbf{\bibinfo{volume}{91}}, \bibinfo{pages}{213103}
  (\bibinfo{year}{2007}).

\bibitem[{\citenamefont{Jones et~al.}(2006)\citenamefont{Jones, Hu, Yang, Yang,
  Hajdaj, and Hehein}}]{Jones06}
\bibinfo{author}{\bibfnamefont{G.~M.} \bibnamefont{Jones}},
  \bibinfo{author}{\bibfnamefont{B.~H.} \bibnamefont{Hu}},
  \bibinfo{author}{\bibfnamefont{C.~H.} \bibnamefont{Yang}},
  \bibinfo{author}{\bibfnamefont{M.~J.} \bibnamefont{Yang}},
  \bibinfo{author}{\bibfnamefont{R.}~\bibnamefont{Hajdaj}}, \bibnamefont{and}
  \bibinfo{author}{\bibfnamefont{G.}~\bibnamefont{Hehein}},
  \bibinfo{journal}{Appl. Phys. Lett.} \textbf{\bibinfo{volume}{89}},
  \bibinfo{pages}{073106} (\bibinfo{year}{2006}).

\bibitem[{\citenamefont{Angus et~al.}(2007)\citenamefont{Angus, Ferguson,
  Dzurak, and Clark}}]{Angus07}
\bibinfo{author}{\bibfnamefont{S.~J.} \bibnamefont{Angus}},
  \bibinfo{author}{\bibfnamefont{A.~J.} \bibnamefont{Ferguson}},
  \bibinfo{author}{\bibfnamefont{A.~S.} \bibnamefont{Dzurak}},
  \bibnamefont{and} \bibinfo{author}{\bibfnamefont{R.~G.} \bibnamefont{Clark}},
  \bibinfo{journal}{Nano. Lett.} \textbf{\bibinfo{volume}{7}},
  \bibinfo{pages}{2051} (\bibinfo{year}{2007}).

\bibitem[{\citenamefont{Kouwenhoven et~al.}(2001)\citenamefont{Kouwenhoven,
  Austing, and Tarucha}}]{Kouwen01}
\bibinfo{author}{\bibfnamefont{L.~P.} \bibnamefont{Kouwenhoven}},
  \bibinfo{author}{\bibfnamefont{D.~G.} \bibnamefont{Austing}},
  \bibnamefont{and} \bibinfo{author}{\bibfnamefont{S.}~\bibnamefont{Tarucha}},
  \bibinfo{journal}{Rep. Prog. Phys.} \textbf{\bibinfo{volume}{64}},
  \bibinfo{pages}{701} (\bibinfo{year}{2001}).

\bibitem[{\citenamefont{Beenakker}(1991)}]{Beena91}
\bibinfo{author}{\bibfnamefont{C.~W.~J.} \bibnamefont{Beenakker}},
  \bibinfo{journal}{Phys. Rev. B} \textbf{\bibinfo{volume}{44}},
  \bibinfo{pages}{1646} (\bibinfo{year}{1991}).

\bibitem[{\citenamefont{van~der Wiel et~al.}(2003)\citenamefont{van~der Wiel,
  Franceschi, Elzerman, Fujisawa, Tarucha, and Kouwenhoven}}]{Van03}
\bibinfo{author}{\bibfnamefont{W.~G.} \bibnamefont{van~der Wiel}},
  \bibinfo{author}{\bibfnamefont{S.~D.} \bibnamefont{Franceschi}},
  \bibinfo{author}{\bibfnamefont{J.~M.} \bibnamefont{Elzerman}},
  \bibinfo{author}{\bibfnamefont{T.}~\bibnamefont{Fujisawa}},
  \bibinfo{author}{\bibfnamefont{S.}~\bibnamefont{Tarucha}}, \bibnamefont{and}
  \bibinfo{author}{\bibfnamefont{L.~P.} \bibnamefont{Kouwenhoven}},
  \bibinfo{journal}{Rev. Mod. Phys.} \textbf{\bibinfo{volume}{75}},
  \bibinfo{pages}{1} (\bibinfo{year}{2003}).

\bibitem[{\citenamefont{Kastner}(1992)}]{Kastner92}
\bibinfo{author}{\bibfnamefont{M.~A.} \bibnamefont{Kastner}},
  \bibinfo{journal}{Rev. Mod. Phys.} \textbf{\bibinfo{volume}{64}},
  \bibinfo{pages}{849} (\bibinfo{year}{1992}).

\bibitem[{\citenamefont{Ishikuro and Hiramoto}(1997)}]{Ishikuro97}
\bibinfo{author}{\bibfnamefont{H.}~\bibnamefont{Ishikuro}} \bibnamefont{and}
  \bibinfo{author}{\bibfnamefont{T.}~\bibnamefont{Hiramoto}},
  \bibinfo{journal}{Appl. Phys. Lett.} \textbf{\bibinfo{volume}{71}},
  \bibinfo{pages}{3691} (\bibinfo{year}{1997}).

\bibitem[{\citenamefont{Lin et~al.}(2007)\citenamefont{Lin, Aravind, Wu, Wu,
  Kuan, Kuo, and Chen}}]{MCLin07}
\bibinfo{author}{\bibfnamefont{M.~C.} \bibnamefont{Lin}},
  \bibinfo{author}{\bibfnamefont{K.}~\bibnamefont{Aravind}},
  \bibinfo{author}{\bibfnamefont{C.~S.} \bibnamefont{Wu}},
  \bibinfo{author}{\bibfnamefont{Y.~P.} \bibnamefont{Wu}},
  \bibinfo{author}{\bibfnamefont{C.~H.} \bibnamefont{Kuan}},
  \bibinfo{author}{\bibfnamefont{W.}~\bibnamefont{Kuo}}, \bibnamefont{and}
  \bibinfo{author}{\bibfnamefont{C.~D.} \bibnamefont{Chen}},
  \bibinfo{journal}{Appl. Phys. Lett.} \textbf{\bibinfo{volume}{90}},
  \bibinfo{pages}{032106} (\bibinfo{year}{2007}).

\bibitem[{\citenamefont{Sellier et~al.}(2006)\citenamefont{Sellier, Lansbergen,
  Caro, Rogge, Collaert, Ferain, Jurczak, and Biesemans}}]{Sellier06}
\bibinfo{author}{\bibfnamefont{H.}~\bibnamefont{Sellier}},
  \bibinfo{author}{\bibfnamefont{G.~P.} \bibnamefont{Lansbergen}},
  \bibinfo{author}{\bibfnamefont{J.}~\bibnamefont{Caro}},
  \bibinfo{author}{\bibfnamefont{S.}~\bibnamefont{Rogge}},
  \bibinfo{author}{\bibfnamefont{N.}~\bibnamefont{Collaert}},
  \bibinfo{author}{\bibfnamefont{I.}~\bibnamefont{Ferain}},
  \bibinfo{author}{\bibfnamefont{M.}~\bibnamefont{Jurczak}}, \bibnamefont{and}
  \bibinfo{author}{\bibfnamefont{S.}~\bibnamefont{Biesemans}},
  \bibinfo{journal}{Phys. Rev. Lett.} \textbf{\bibinfo{volume}{97}},
  \bibinfo{pages}{206805} (\bibinfo{year}{2006}).

\bibitem[{\citenamefont{Rokhinson et~al.}(2001)\citenamefont{Rokhinson, Guo,
  Chou, and Tsui}}]{Rohkin01}
\bibinfo{author}{\bibfnamefont{L.~P.} \bibnamefont{Rokhinson}},
  \bibinfo{author}{\bibfnamefont{L.~J.} \bibnamefont{Guo}},
  \bibinfo{author}{\bibfnamefont{S.~Y.} \bibnamefont{Chou}}, \bibnamefont{and}
  \bibinfo{author}{\bibfnamefont{D.~C.} \bibnamefont{Tsui}},
  \bibinfo{journal}{Phys. Rev. B} \textbf{\bibinfo{volume}{63}},
  \bibinfo{pages}{035321} (\bibinfo{year}{2001}).

\bibitem[{\citenamefont{Ando et~al.}(1982)\citenamefont{Ando, Fowler, and
  Stern}}]{Ando82}
\bibinfo{author}{\bibfnamefont{T.}~\bibnamefont{Ando}},
  \bibinfo{author}{\bibfnamefont{A.~B.} \bibnamefont{Fowler}},
  \bibnamefont{and} \bibinfo{author}{\bibfnamefont{F.}~\bibnamefont{Stern}},
  \bibinfo{journal}{Rev. Mod. Phys.} \textbf{\bibinfo{volume}{54}},
  \bibinfo{pages}{437} (\bibinfo{year}{1982}).

\bibitem[{\citenamefont{Goswami et~al.}(2007)\citenamefont{Goswami, Slinker,
  Friesen, McGuire, Truitt, Tahan, Klein, Chu, Mooney, van~der Weide
  et~al.}}]{Srijit07}
\bibinfo{author}{\bibfnamefont{S.}~\bibnamefont{Goswami}},
  \bibinfo{author}{\bibfnamefont{K.~A.} \bibnamefont{Slinker}},
  \bibinfo{author}{\bibfnamefont{M.}~\bibnamefont{Friesen}},
  \bibinfo{author}{\bibfnamefont{L.~M.} \bibnamefont{McGuire}},
  \bibinfo{author}{\bibfnamefont{J.~L.} \bibnamefont{Truitt}},
  \bibinfo{author}{\bibfnamefont{C.}~\bibnamefont{Tahan}},
  \bibinfo{author}{\bibfnamefont{L.~J.} \bibnamefont{Klein}},
  \bibinfo{author}{\bibfnamefont{J.~O.} \bibnamefont{Chu}},
  \bibinfo{author}{\bibfnamefont{P.~M.} \bibnamefont{Mooney}},
  \bibinfo{author}{\bibfnamefont{D.~W.} \bibnamefont{van~der Weide}},
  \bibnamefont{et~al.}, \bibinfo{journal}{Nature Physics}
  \textbf{\bibinfo{volume}{3}}, \bibinfo{pages}{41} (\bibinfo{year}{2007}).

\bibitem[{\citenamefont{Takashina et~al.}(2006)\citenamefont{Takashina, Ono,
  Fujiwara, Takahashi, and Hirayama}}]{Taka06}
\bibinfo{author}{\bibfnamefont{K.}~\bibnamefont{Takashina}},
  \bibinfo{author}{\bibfnamefont{Y.}~\bibnamefont{Ono}},
  \bibinfo{author}{\bibfnamefont{A.}~\bibnamefont{Fujiwara}},
  \bibinfo{author}{\bibfnamefont{Y.}~\bibnamefont{Takahashi}},
  \bibnamefont{and} \bibinfo{author}{\bibfnamefont{Y.}~\bibnamefont{Hirayama}},
  \bibinfo{journal}{Phys. Rev. Lett.} \textbf{\bibinfo{volume}{96}},
  \bibinfo{pages}{236801} (\bibinfo{year}{2006}).

\bibitem[{\citenamefont{Liu et~al.}(2008)\citenamefont{Liu, Fujisawa, Ono,
  Inokawa, Fujiwara, Takashina, and Hirayama}}]{Liu08}
\bibinfo{author}{\bibfnamefont{H.~W.} \bibnamefont{Liu}},
  \bibinfo{author}{\bibfnamefont{T.}~\bibnamefont{Fujisawa}},
  \bibinfo{author}{\bibfnamefont{Y.}~\bibnamefont{Ono}},
  \bibinfo{author}{\bibfnamefont{H.}~\bibnamefont{Inokawa}},
  \bibinfo{author}{\bibfnamefont{A.}~\bibnamefont{Fujiwara}},
  \bibinfo{author}{\bibfnamefont{K.}~\bibnamefont{Takashina}},
  \bibnamefont{and} \bibinfo{author}{\bibfnamefont{Y.}~\bibnamefont{Hirayama}},
  \bibinfo{journal}{Phys. Rev. B} \textbf{\bibinfo{volume}{77}},
  \bibinfo{pages}{073310} (\bibinfo{year}{2008}).

\bibitem[{\citenamefont{Ciorga et~al.}(2000)\citenamefont{Ciorga, Sachrajda,
  Hawrylaka, Gould, Zawadzki, Jullian, Feng, and Wasilewski}}]{Ciorga00}
\bibinfo{author}{\bibfnamefont{M.}~\bibnamefont{Ciorga}},
  \bibinfo{author}{\bibfnamefont{A.~S.} \bibnamefont{Sachrajda}},
  \bibinfo{author}{\bibfnamefont{P.}~\bibnamefont{Hawrylaka}},
  \bibinfo{author}{\bibfnamefont{C.}~\bibnamefont{Gould}},
  \bibinfo{author}{\bibfnamefont{P.}~\bibnamefont{Zawadzki}},
  \bibinfo{author}{\bibfnamefont{S.}~\bibnamefont{Jullian}},
  \bibinfo{author}{\bibfnamefont{Y.}~\bibnamefont{Feng}}, \bibnamefont{and}
  \bibinfo{author}{\bibfnamefont{Z.}~\bibnamefont{Wasilewski}},
  \bibinfo{journal}{Phys. Rev. B} \textbf{\bibinfo{volume}{61}},
  \bibinfo{pages}{16315} (\bibinfo{year}{2000}).

\bibitem[{\citenamefont{Vitkalov et~al.}(2001)\citenamefont{Vitkalov, Sarachik,
  and Klapwijk}}]{Vitka01}
\bibinfo{author}{\bibfnamefont{S.~A.} \bibnamefont{Vitkalov}},
  \bibinfo{author}{\bibfnamefont{M.~P.} \bibnamefont{Sarachik}},
  \bibnamefont{and} \bibinfo{author}{\bibfnamefont{T.~M.}
  \bibnamefont{Klapwijk}}, \bibinfo{journal}{Phys. Rev. B}
  \textbf{\bibinfo{volume}{64}}, \bibinfo{pages}{073101}
  (\bibinfo{year}{2001}).

\bibitem[{\citenamefont{Shashkin et~al.}(2001)\citenamefont{Shashkin,
  Kravchenko, Dolgopolov, and Klapwijk}}]{Shash01}
\bibinfo{author}{\bibfnamefont{A.~A.} \bibnamefont{Shashkin}},
  \bibinfo{author}{\bibfnamefont{S.~V.} \bibnamefont{Kravchenko}},
  \bibinfo{author}{\bibfnamefont{V.~T.} \bibnamefont{Dolgopolov}},
  \bibnamefont{and} \bibinfo{author}{\bibfnamefont{T.~M.}
  \bibnamefont{Klapwijk}}, \bibinfo{journal}{Phys. Rev. Lett.}
  \textbf{\bibinfo{volume}{87}}, \bibinfo{pages}{086801}
  (\bibinfo{year}{2001}).

\bibitem[{\citenamefont{Sachrajda et~al.}(2001)\citenamefont{Sachrajda,
  Hawrylaka, Ciorga, Gould, and Zawadzki}}]{Sach01}
\bibinfo{author}{\bibfnamefont{A.~S.} \bibnamefont{Sachrajda}},
  \bibinfo{author}{\bibfnamefont{P.}~\bibnamefont{Hawrylaka}},
  \bibinfo{author}{\bibfnamefont{M.}~\bibnamefont{Ciorga}},
  \bibinfo{author}{\bibfnamefont{C.}~\bibnamefont{Gould}}, \bibnamefont{and}
  \bibinfo{author}{\bibfnamefont{P.}~\bibnamefont{Zawadzki}},
  \bibinfo{journal}{Physica E} \textbf{\bibinfo{volume}{10}},
  \bibinfo{pages}{493} (\bibinfo{year}{2001}).

\end{thebibliography}

\end{document}